\documentclass[aps,prl,twocolumn,showpacs,floatfix,superscriptaddress]{revtex4-1}
\usepackage{graphicx,amsfonts,amssymb,amsmath,hyperref}
\usepackage{textcomp}
\usepackage{esint}
\usepackage{bbold}
\usepackage{braket}
\usepackage{amsmath}
\usepackage{amsfonts} 
\newif\ifhyper
\hypertrue
\ifhyper
\hypersetup{
  citecolor = {green},
  colorlinks = {true}, 
  urlcolor = {blue} 
}

\newcommand{\Tr}{{\rm Tr}} 
 
\newcommand{\re}{{\rm Re}} 
\newcommand{\im}{{\rm Im}} 
 
\newcommand{\mean}[1]{\langle #1 \rangle}

\newcommand{\rmd}{{\rm d}}

\def\half{\frac{1}{2}}

\def\w{\omega}

\def\dt{\partial_t}

\def\dt{\partial_t}
\def\calA{{\cal A}}
\def\calB{{\cal B}}

\def\calJ{{\cal J}}
\def\calL{{\cal L}} 
\def\calM{{\cal M}}

\def\calP{{\cal P}}  
 
\def\calS{{\cal S}}
\def\calU{{\cal U}}
\def\calZ{{\cal Z}}

\begin{document}

\title{Exact dynamics and thermalization of an open bosonic quantum system in presence of a quantum phase transition induced by the environment} 

\author{A. Ran\c{c}on}
\affiliation{James Franck Institute and Department of Physics, University of Chicago, Chicago, Illinois 60637, USA}
\author{J. Bonart}
\affiliation{Laboratoire de Physique Th\'eorique et Hautes Energies, Universit\'e Pierre et Marie Curie -- Paris VI - 4 Place Jussieu, 75252 Paris Cedex 05, France}
\date{\today}
 
\begin{abstract} 
We derive the exact out-of-equilibrium Wigner function of a bosonic mode linearly coupled to a bosonic bath of arbitrary spectral density. Our solution does not rely on any master equation approach and it therefore also correctly describes a bosonic mode which is initially entangled with its environment. It has been recently suggested that non-Markovian quantum effects lead to dissipationless dynamics in the case of a strong coupling to a bath whose spectral density has a support bounded from below. We show in this work that such a system undergoes a quantum phase transition at some critical bath coupling strength. The apparent dissipationless dynamics then correspond to the relaxation towards the new ground-state.  
\end{abstract}

\pacs{03.65.Yz, 03.65.Ta, 05.70.Ln}
\maketitle

{\it Introduction. -} 
The decoherence of an open quantum system has been studied by a variety of methods in the past~\cite{BreuerBook}. One frequently used approach is based on an exact (non-Markovian) master equation for the reduced density-matrix (RDM) of the system~\cite{Hu1992,Ford2001,Fleming2011,Lei2011}.
While its success  is undeniable, all known master equations are however derived under the assumptions that the system and its environment are initially uncoupled. The question of how to describe a non-Markovian time evolution of open quantum systems, preferably \emph{via} a closed equation for the RDM subject to \emph{arbitrary} [i.e. entangled] initial conditions, is still under debate.


The objective of this letter is twofold. We first show how to derive the exact Wigner function of a bosonic mode linearly coupled to a bosonic bath in the general case of arbitrary, possibly \emph{non-factorizing} initial conditions. Hence, our analysis will go beyond the range of validity of the master equations. We then focus on a particular environment which conserves the total number of excitations. We revisit the issue concerning ``dissipationless dynamics'' recently raised in the literature~\cite{Lei2011,ZhangWM2012}, where it has been argued that the relaxation is inhibited by strong non-Markovian effects in the bath. We show that this peculiar behavior is on the contrary related to a [static] quantum phase transition in the global system and hence it does not directly result from dynamic memory effects in the environment. 


\vspace{1mm}
{\it Model. -} 
Let us consider the dynamics of a bosonic $(\hat a^\dagger,\hat a)$-mode [of mass $m$ and frequency $\w_0$], the ``system'' $\calS$, in contact with a ``bath'' $\calB$ modeled as a set of $N_\calB$ harmonic oscillators $(\hat b_i^\dagger,\hat b_i)$ [of mass $m_i$ and frequency $\w_i$]. The Hamiltonian is then given by [$\hbar=k_B=1$ throughout this letter]
\begin{equation}\label{eq:Hinit}
\hat H=\w_0\,\hat a^\dag\, \hat a +\hat H_\calB+\hat H_c \; ,
\end{equation}
where $\hat H_\calB=\sum_{i=1}^{N_\calB} \w_i\, \hat b^\dag_i\, \hat b_i$ is the Hamiltonian of the bath and $\hat H_c$ describes the $\calS\calB$-coupling. We assume $\hat H_c$ to be quadratic in the ladder operators. Furthermore, all the parameters appearing in $\hat H$ [and $\hat H_c$] can depend on time, implying that the whole system $\calU=\calS\oplus\calB$ may be driven out-of-equilibrium. The general dynamics after \emph{arbitrary initial conditions} governed by Eq.~(\ref{eq:Hinit}) will be analyzed in the following in terms of the \emph{Wigner function} $W_t(q,p)$ of the RDM $\hat \rho$ of the system: $\hat \rho(t)=\Tr_\calB\Big\{\hat\rho_\calU(t)\Big\}$, with $\hat\rho_\calU(t)=\hat U(t)\hat\rho_\calU(0)\hat U^\dagger(t)$ the total density-matrix at time $t$ with initial condition $\hat\rho_\calU(0)$ [the evolution operator $\hat U(t)$ solves $i\dt \hat U(t)=\hat H(t)\hat U(t)$]. $W_t(q,p)$ is defined as 
\begin{equation}
W_t(q,p)=\frac{1}{2\pi}\int \rmd u \ e^{-i p u}\bra{q +\frac{u}{2}}\hat \rho(t)\ket{q -\frac{u}{2}} \; ,
\label{eq_def_W}
\end{equation}
with $|q\pm u/2\rangle$ position eigenstates. 

In the last part of this letter we discuss the effects of a particular environment, the so-called \emph{resonant} bath for which 
\begin{equation}\label{eq:Hrmr}
  \hat H_c= \hat H_{\rm r}=\sum_i C_i (\hat a\ \hat b_i^\dag+\hat a^\dag\, \hat b_i) \; .
\end{equation}
Let us first analyze the physical content of the above model. The canonical model for dissipative quantum dynamics is the \emph{quantum Brownian motion} (QBM) modeled by 
$\hat H_c= \hat H_{\rm QBM}=\sum_i C_i (\hat a+\hat a^\dag)(\hat b_i+\hat b_i^\dag)$. Under certain circumstances one disregards the term $\hat a\ \hat b_i+\hat a^\dag\hat b_i^\dag$ which then leads to $\hat H_{\rm r}$ defined in Eq.~(\ref{eq:Hrmr}). This is the well-known rotating wave approximation (RWA). Note however, that the Eq.~(\ref{eq:Hrmr}) can be more than a simple approximation of QBM since it models an interaction which conserves the total number of particles $\hat N=\hat a^\dag \,\hat a +\sum_i\hat b_i^\dag\,\hat b_i$, a symmetry that can be imposed by nature right from the start to describe possible experimental setups~\cite{ref_exp}. 

Note that the influence of the bath on the system in the resonant model is completely determined by the spectral function $S(\w)=2\pi\sum_i C_i^2 \delta(\w-\w_i)$. In the present case the spectral function is non-zero only for $\w\ge 0$ [in contrast to QBM]. If the number of bath modes is infinite we can also write 
$S(\w)=\eta\, \w^s f(\w/\w_c)$, where $\eta\propto C_i^2$ characterizes the strength of the $\calS\calB$-coupling and $f(x)$ is a cut-off function. Depending on the exponent $s$ which describes the low-$\w$ behavior of $S(\w)$ the bath is said to be Ohmic, super-Ohmic or sub-Ohmic [$s=1$, $s>1$ or $s<1$, respectively].

\vspace{1mm}
{\it Wigner function of a Gaussian initial state. -} 
In order to obtain  the exact Wigner function of $\calS$, we analyze first  $\hat \rho^G(t)$. The superscript $G$ indicates that we start from a Gaussian initial condition. One can show that the matrix elements $\rho^G_{x,y}(t)=\bra x \hat \rho^G(t) \ket y$ satisfy [see Appendix]
\begin{equation}
\rho^G_{x,y}(t)=\int \rmd r \ e^{-\frac{i}{2}r(x+y)}\mean{e^{-i(x-y)\hat p+ir\hat q}}_t \; ,
\label{eq_rhoxy}
\end{equation}
where $\mean{\cdots}_t=\Tr_\calU\Big\{\hat \rho_\calU(t)\cdots\Big\}$ is the time-dependent statistical average and $\hat q=\frac{1}{\sqrt{2m\w_0}}(\hat a^\dag +\hat a)$, $\hat p=i\sqrt{\frac{m\w_0}{2}}(\hat a^\dag -\hat a)$ are the position and momentum operators of $\calS$. Since we assume in this paragraph $\rho_\calU(0)$ to be Gaussian, the average in  Eq.~\eqref{eq_rhoxy} can be readily done by realizing that -- within a path-integral formalism~\cite{Bonart2012} -- $\hat p$ and $\hat q$ become Gaussian random variables. One then obtains
\begin{align}
\label{eq_mean}
&\mean{e^{-i(x-y)\hat p+ir\hat q}}_t = \\
&\qquad\qquad e^{i r \bar q-\frac{C_{pp}}{2}(x-y)^2-\frac{C_{qq}}{2}r^2+r(x-y)C_{qp}-i\bar p(x-y)} \; ,
\nonumber
\end{align}
where we denote the mean value by $\bar A(t)=\mean{\hat A}_t$ and the correlation functions by $C_{AB}(t)=\half\mean{\hat A\,\hat B+\hat B\,\hat A}_t-\bar A(t)\bar B(t)$ for any two operators $\hat A$ and $\hat B$. These correlation functions can be determined by using path integral methods~\cite{Grabert1988} or by averaging the solution of the equations of motion over the initial density-matrix $\rho_\calU(0)$ [see a more detailed discussion below].

From Eq.~\eqref{eq_rhoxy} we find in conjunction with Eq.~(\ref{eq_mean}) the final expression for the RDM of a Gaussian initial condition after the time evolution under a quadratic [possibly time-dependent] Hamiltonian:
\begin{equation}
\rho^G_{x,y}(t)=\frac{e^{\tilde m(x-\bar q)(y-\bar q)-\frac{m}{2}(x-\bar q)^2-\frac{m^*}{2}(y-\bar q)^2-i\bar p(x-y)}}{\sqrt{2\pi C_{xx}}} \; ,
\label{eq_rhofin}
\end{equation}
with $\tilde m=C_{pp}-(\frac{1}{4}+C_{xp}^2)/C_{xx}$ and $m=C_{pp}+(\half-i C_{xp})^2/C_{xx}$. Note that this result also allows the study of quantum quenches by setting $\rho_\calU(0)=e^{-\hat H_{\rm init}/T}/\Tr \big\{e^{- \hat H_{\rm init}/T}\big\}$ with $\hat H_{\rm init}$, a quadratic [possibly interacting] Hamiltonian different from $\hat H$, and $T$,  the initial temperature of the system.

Let us now determine the Wigner function associated with the Gaussian density-matrix~(\ref{eq_rhofin}) by using the definition~(\ref{eq_def_W}). After a straightforward calculation one finds
\begin{equation}
W_t^G(z)=\frac{e^{-\half (z-\bar z)^T\cdot{\calA_t^G}^{-1}\cdot(z-\bar z)}}{2\pi \sqrt{\det \calA^G_t}} \;.
\label{eq_W_G}
\end{equation}
We introduced the vector notation $z^T=(q,p)$ and $\zeta_i=(q_i,p_i)$ [which we shall use later] for the mode $i$ of $\calB$, as well as the euclidean scalar product $\cdot$ of $\mathbb{R}^2$. 
Furthermore, $\calA^G_t=\half\mean{\hat z\cdot\hat z^T+(\hat z\cdot\hat z^T)^T}_t^G-\bar z(t)\cdot\bar z(t)^T$ is the covariance matrix [note that $\hat z\cdot\hat z^T\neq(\hat z\cdot\hat z^T)^T$ since $\hat q$ and $\hat p$ do not commute] which can be recast as
\begin{equation}
\calA^G_t=\begin{pmatrix}
C_{qq}(t) & C_{qp}(t) \\ C_{qp}(t) & C_{pp}(t)
\end{pmatrix} \; .
\end{equation}
The various correlators can be computed by using the equations of motion of the operators $\hat z$ and $\hat \zeta_i$: 
\begin{align}
i\dt \hat z(t)&=M_0(t)\cdot\hat z(t)+\sum_i \tilde M_{0i}(t)\cdot\hat \zeta_i(t),\\
i \dt \hat \zeta_i(t)& =  M_i(t)\cdot\hat \zeta_i(t)+\tilde M_{i0}(t)\cdot\hat z(t)\;,
\end{align}
where $M_0$, $M_i$, $\tilde M_{0i}$, and $\tilde M_{i0}$ are two-dimensional matrices, the details of which depend on the model. By solving first the equations of the bath operators we find the solutions $\hat \zeta_i(t)=\calL_i(t)\cdot\hat \zeta_i(0)+\int_0^t \rmd\tau\calL_i(t-\tau)\cdot\tilde M_{i0}(\tau)\cdot\hat z(\tau)$, with $i\dt \calL_i(t)=M_i(t)\cdot\calL_i(t)$ and $\calL_i(0)=1$. By inserting these solutions into the equations of motion of $\hat z$ we further obtain
\begin{align}
\hat z(t)=\Phi(t)\cdot\hat z(0)+\sum_i \calM_i(t)\cdot\hat \zeta_i(0) \; ,
\label{eq_motion}
\end{align}
with 
\begin{equation}
\calM_i(t)=\int_0^t \rmd\tau\ \Phi(t-\tau)\cdot\tilde M_{0i}(\tau)\cdot\calL_i(\tau)\;,
\label{eq_Mi}
\end{equation}
and $\Phi(t)$ the solution of 
\begin{align}
0 &= i\dt\Phi(t)-M_0(t)\cdot\Phi(t)\\
&-\sum_i \int_0^t \rmd\tau \tilde M_{0i}(t)\cdot\calL_i(t-\tau)\cdot\tilde M_{i0}(\tau)\cdot\Phi(\tau)\;,\nonumber 
\end{align}
with $\Phi(0)=1$. From Eq.~(\ref{eq_motion}) the covariance matrix $\calA^G_t$ can be readily computed by averaging over $\hat\rho_\calU(0)$ \cite{note0}.

\vspace{1mm}
{\it Wigner function of an arbitrary initial state. -} 
We show now how the density-matrix of an arbitrary non-Gaussian initial condition can be constructed from the density-matrix of a coherent-state initial condition. Note that coherent-states are Gaussian states such that we can make the link with the previous section. 

It is well known that any density-matrix can be written as a diagonal matrix in the coherent-state basis by using the Glauber-Sudarshan $P$-function [see \emph{e.g.}~\cite{LeonhardtBook}]: 
$\hat\rho_\calU(0)=\int \rmd^2\alpha \prod_i d^2\beta_i P_0\big(\alpha;\{\beta_i\}\big) \ket{\alpha;\{\beta_i\}}\bra{\alpha;\{\beta_i\}}$, where a state of $\calU$ is written as $\ket{\alpha;\{\beta_i\}}$ with $\alpha$ the state of $\calS$ and $\{ \beta_i\}=\{\beta_1,\beta_2,\cdots\}$ the state of $\calB$. By taking a partial trace over the $\calB$-states we find the RDM 
\begin{equation}
\rho_{x,y}(t)=\int \rmd^2\alpha \prod_i \rmd^2\beta_i \ P_0(\alpha;\{\beta_i\})\  \rho^G_{x,y}(t) \; ,
\label{eq_arbitrary}
\end{equation}
where $ \rho^G_{x,y}(t)$ is now associated with the \emph{particular} Gaussian initial condition $\hat\rho^{\ket{\alpha;\beta}} = \ket{\alpha;\{\beta_i\}}\bra{\alpha;\{\beta_i\}}$. 


Hence, by using Eqs.~\eqref{eq_arbitrary} and~\eqref{eq_def_W} the Wigner function corresponding to an arbitrary initial state is given by
\begin{equation}
W_t(z)=\int \rmd\tilde z \prod_i \rmd\tilde \zeta_i \ W_t^G(z) \ P_0(\tilde z,\tilde \zeta_i) \; ,
\label{eq_Wf_int}
\end{equation}
where $W_t^G(z)$ is given in Eq.~(\ref{eq_W_G}) with the initial condition $\hat\rho^{\ket{\alpha;\beta}}$. Note that the $P$-function contains all the information on the non-Gaussian initial condition.
However, the $P$-function is in general highly singular and therefore not suited for concrete applications. We therefore proceed by eliminating $P_0(\tilde z,\tilde \zeta_i)$. 
Let us first express the problem solely in terms of the variables $\tilde z$ and $\{\tilde \zeta_i\}$. By definition, we can write $\int \rmd\tilde z \prod_i \rmd\tilde \zeta_i  P_0(\tilde z,\tilde \zeta_i)=\rmd^2\alpha \prod_i \rmd^2\beta_i P_0(\alpha;\{\beta_i\})$ with $\tilde z^T=(\sqrt{2/m\w_0} \ \re(\alpha), \sqrt{2m\w_0} \ \im(\alpha))$ and similar relations between the $\tilde \zeta_i$ and $\beta_i$. Hence, $W_t^G(z)$ now depends \emph{via} its initial condition $\hat\rho^{\ket{\alpha;\beta}}=\ket{\tilde z;\{\tilde \zeta_i\}}\bra{\tilde z;\{\tilde \zeta_i\}}$ on the new variables $\tilde z$ and $\tilde \zeta_i$, too. Note that we have relabeled the coherent state $\ket{\alpha;\{\beta_i\}}$ by virtue of the above relation between $\tilde z^T$, $\re ( \alpha)$ and $\im ( \alpha)$. 
In the following we use the convention for the Fourier transform $\tilde F(k) = \int{\rm d}z \ e^{-ik^T\cdot z}F(z)$ of a function $F(z)$, defining analogously the Fourier transform of functions of many variables.

Second, let us introduce the Wigner function $W_0(\tilde z,\{\tilde \zeta_i\})$ of the [non-Gaussian] initial condition $\hat \rho_\calU(0)$. According to~\cite{LeonhardtBook} it can be written in terms of $P_0(\tilde z,\tilde\zeta_i)$ in the Fourier domain as
 \begin{equation}
\hspace{-0.2cm}\tilde W_0(k;\{\kappa_i\})=\tilde P_0(k;\{\kappa_i\})\ e^{-\half k^T\cdot\calA_0\cdot k - \sum_i\half \kappa_i^T\cdot\calA_i\cdot\kappa_i} \;,
\label{eq_SM_WP_fourier}
\end{equation}
%
%
%
where 
\begin{equation}
\calA_0 = \begin{pmatrix}\frac{1}{2m\w_0} & 0 \\ 0 &\frac{m\w_0}{2} \end{pmatrix}\;,\;\;
\calA_i = \begin{pmatrix}\frac{1}{2m_i\w_i} & 0 \\ 0 &\frac{m_i\w_i}{2}\end{pmatrix}\;,
\label{eq_SM_calA}	
\end{equation}
are the covariance matrices of a coherent state.  
 %
%

Third, the Fourier transform of Eq.~\eqref{eq_Wf_int} with respect to $z$ [using Eq.~\eqref{eq_W_G}]
now reads
\begin{equation}
\hspace{-0.2cm}\tilde W_t(k)=\int \rmd\tilde z \prod_i d\tilde \zeta_i \ e^{-\half k^T\cdot\calA^G_t\cdot k-ik^T\cdot\bar z} \ P_0(\tilde z,\{\tilde \zeta_i\}) \; .
\label{eq:SM_Wf_fourier}
\end{equation}
From Eq.~\eqref{eq_motion} one easily shows that 
$
\bar z(t)=\Phi(t)\cdot\tilde z+\sum_i\calM_i(t)\cdot\tilde \zeta_i\;,
$
and Eq.~\eqref{eq:SM_Wf_fourier} can be recast as
\begin{align}
\tilde W_t(k) = &e^{-\half k^T\cdot\left[\calA^G_t-\Phi\cdot\calA_0\cdot\Phi^T-\sum_i\calM_i\cdot\calA_i\cdot\calM_i^T\right]\cdot k} \nonumber\\
&\times \tilde W_0( \Phi^T\cdot k,\{\calM_i^T\cdot k\}) \; .
\label{eq:SM_Wf_fourier2}
\end{align}
Finally, with the definition $\calA^G_t=\half\mean{\hat z\cdot\hat z^T+(\hat z\cdot\hat z^T)^T}_t^G-\bar z(t)\cdot\bar z(t)^T$ and by noting that $\calA^G_t$ has to be calculated with $\hat\rho^{\ket{\alpha;\beta}}$ [see above] one can show after some algebra that $\calA_t=\Phi\cdot\calA_0\cdot\Phi^T-\sum_i\calM_i\cdot\calA_i\cdot\calM_i^T$: The exponent in the rhs of Eq.~(\ref{eq:SM_Wf_fourier2}) thus cancels out exactly and
\begin{equation}
W_t(z)=\int \frac{\rmd k}{2\pi}\ e^{i k^T\cdot z} \ \tilde W_0\left(\Phi^T(t)\cdot k,\{\calM_i^T(t)\cdot k\}\right) \; .
\label{eq_Wf}
\end{equation}
Note that Eq.~(\ref{eq_Wf}) holds regardless of the entanglement in the initial condition \cite{note1}. In the particular case where the initial condition is factorized, $\hat \rho_\calU(0)=\hat\rho_\calS\otimes \hat \rho_\calB$, we further have $W_0\big(z,\{\zeta_i\}\big)=W_\calS(z)\ W_\calB\big(\{\zeta_i\}\big)$. By defining the \emph{propagator} $K_t(z)=\int \frac{\rmd k}{2\pi} \ e^{ik^T\cdot z}\tilde W_\calB\big(\{\calM_i^T(t)\cdot k\}\big)$ we can write
\begin{equation}
W_t(z)=\int \rmd\tilde z\  K_t\left(z-\Phi(t)\cdot\tilde z\right)\ W_\calS(\tilde z) \; .
\label{eq_factorized}
\end{equation}
Since $K_{t=0}(z)=\delta(z)$ [by definition $\calM_i(0)=0$ and by normalization $\tilde W_\calB(\{\kappa_i=0\})=1$], 
 the term ``propagator'' is particularly well suited for $K_t$.

Eq.~(\ref{eq_Wf}) and in particular Eq.~(\ref{eq_factorized}) will be used in the following to analyze the relaxation dynamics of the resonant model.

\vspace{1mm}
{\it Quantum phase transition and equilibration in the resonant model. -} 
Let us come back to the resonant model with time-independent coupling. The central aspect of our argumentation is the analysis of the time evolution of $\hat \rho_\calU(0)=\ket{1;\{0\}}\bra{1;\{0\}}$ which describes a factorizing initial condition between the system in its first Fock state and the bath ground-state [\emph{i.e.} at zero temperature]. The equation of motion (\ref{eq_motion}) is now readily solved and [at zero $T$] its solution is totally determined by the Green function $\Phi(t)$ \cite{ZhangWM2012}. More precisely, the dynamics is given by [see below] $u(t) \equiv \Phi_{11}(t) - im\omega_0\Phi_{12}(t)$. We have in the Laplace domain
$
\tilde u(\lambda) = \left[\lambda + i\omega_0 + \Sigma(\lambda)\right]^{-1}
$, 
with the self-energy $\Sigma(\lambda)=\int_0^\infty\frac{{\rm d}\w}{2\pi} S(\w)/(\lambda + i\w)$. The relaxation of the system is then completely determined by the long-time behavior of $u(t)$. 
Note that $\Sigma(\lambda)$ has a branch cut on the imaginary half-axis for $\im\lambda <0$. Generically, for sufficiently strong interactions, $\eta\geq \eta_c$, where $\eta_c$ depends on the details of $S(\w)$, an isolated pole $\lambda_1$ appears in the denominator of $\tilde u(\lambda)$ with $\re\lambda_1=0$ and $\im\lambda_1 > 0$. This pole is defined by the equation $\lambda_1+i\w_0+\Sigma(\lambda_1)=0$. 
In real time such a pole gives rise to a purely oscillatory mode. One thus has $u(t)=\calZ_1 e^{\lambda_1t} + \cdots$ in the long time limit where the ellipsis stands for decaying terms. Here, $\calZ_1=\left[1+\Sigma'(\lambda_1)\right]^{-1}$ is the residue of the pole [see Ref.~\cite{ZhangWM2012} for a detailed discussion]. Accordingly, for $\eta>\eta_c$ it has been argued that the system's relaxation is inhibited by the emergence of this isolated pole and the resulting dynamics have been called ``dissipationless'' in the recent literature~\cite{Lei2011,ZhangWM2012}.

Let us further interpret these formal equations by analyzing the spectrum of $\hat H$. Consider in particular the eigenstates of $\hat H$ with zero and one total excitation (the number of which is conserved), which we write as $\ket{\phi_0}=\ket{0;\{0\}}$ and $\ket{\phi_1}=c_0\ket{1;\{0\}}+\sum_i c_i \ket{0;i}$, where $\ket{0;i}=\hat b^\dag_i\ket{0;\{0\}}$. The vacuum state $\ket{0;\{0\}}$ has zero energy, $e_0=0$, regardless of the $\calS\calB$-coupling. When $\eta$ is very small the vacuum is obviously the ground state of $\mathcal U$. Let us denote the energies of the one-excitation eigenstates by $e_1^{(j)}$: By construction one has $\hat H \ket{\phi_1^j}=e_1^{(j)}\ket{\phi_1^j}$. Let us further denote the smallest of these energies by $e_1 = \min(e_1^{(j)})$ which satisfies [with all the other $e_1^{(j)}$] the condition
\begin{equation}\label{eq:e_1}
e_1=\w_0+\sum_{i=1}^{N_\calB} \frac{C_i^2}{e_1-\w_i}\;.
\end{equation}
Also, $c_0$ is determined by
\begin{equation}\label{eq:c02}
c_0^2=\left(1+\sum_{i=1}^{N_\calB}\frac{C_i^2}{(e_1-\w_i)^2}\right)^{-1}=1-\sum_{i=1}^{N_\calB} c_i^2 \;. 
\end{equation}
The last equation is a consequence of the normalization condition $1=\langle\phi_1\ket{\phi_1}$. Moreover, upon inspecting Eqs.~\eqref{eq:e_1} and (\ref{eq:c02})  it is clear that 
\begin{equation}\label{eq:c02Z}
 \lambda_1 = -i e_1\; ,\;\; c_0^2 = \frac{1}{1 + \Sigma'(e_1)} = \mathcal Z_1\;. 
\end{equation}
Obviously, when $e_1>0$ the sum in Eq.~(\ref{eq:c02}) diverges in the limit of an infinite bath [$N_\calB\to\infty$] and $c_0 \sim 1/\sqrt{N_\calB} \to 0$. However, if Eq.~(\ref{eq:e_1}) has a solution $e_1<0$ then $c_0$ remains finite which implies that $\bra{\phi_1}\hat a^\dag \hat a\ket{\phi_1}=c_0^2>0$ \cite{Gaveau1995}. Moreover, in that case we have $e_1<e_0$ and the ground-state changes due to a level crossing. Such a behavior implies a quantum phase transition \cite{note2}. The critical value $\eta_c$ for this to happen is found by setting $e_1 = 0$ in Eq.~(\ref{eq:e_1}) which translates into $\Sigma(0)=-i\w_0$.

The one-excitation eigensubspace of $\hat H$ is spanned by the states $\ket{\phi_1},\ket{\phi^1_1},\ket{\phi^2_1},\cdots $ and for $\eta>\eta_c$ it is straightforward to show that $e_1<0<e_1^{(j)}$. Let us now make the connection to the system which is initially described by $\hat \rho_\calU(0)=\ket{1;\{0\}}\bra{1;\{0\}}$. In order to find the time evolution we expand $\ket{1;\{0\}}$ into a sum over $|\phi_1\rangle$ and all $\ket{\phi_1^j}$: $\ket{1;\{0\}} = d|\phi_1\rangle + \sum_{j=2}^{N_\calB} d_j\ket{\phi_1^j}$, 
from which we find $d_j = c_0^{(j)}$ by multiplying the previous equation by $\ket{\phi_1^j}$. Hence, after a time lag $t$ 
\begin{align}
  \langle 1 |\hat\rho(t) |1\rangle &= d^2\langle 1,\{0\}|\phi_1\rangle
\langle\phi_1|1,\{0\}\rangle \nonumber\\
&+ \sum_{j=2}^{N_\calB} d_j^2\langle 1,\{0\}|\phi_1\rangle
\langle\phi_1|1,\{0\}\rangle
+ \mathrm{osc} \; ,
\end{align}
where ``$\mathrm{osc}$'' stands for oscillating terms $\sim e^{i t (e_1^{(j)} - e_1^{(j')})}$, $j\ne j'$, which cancel out for $t \to \infty$. 
Since $d_j = c_0^{(j)} = \langle\phi_1|1,\{0\}\rangle \sim 1/\sqrt{N_\calB}$ [note that $e_1^{(j)}>0$] and $d = c_0$ we have in the $N_\calB\to\infty$ limit $\langle 1 |\hat\rho(t) |1\rangle \simeq c_0^2|\langle 1,\{0\}|\phi_1\rangle|^2 = c_0^4$. Accordingly, the RDM relaxes towards 
\begin{equation}\label{eq:123}
  \hat\rho(t\to\infty) = (1-c_0^4) |0\rangle\langle 0| + c_0^4 |1\rangle\langle 1|\; .
\end{equation}
%
%
%
This result can indeed be directly derived within our formalism [see, i.e., Eq.~(\ref{eq_factorized})], as it can be shown that for the initial condition $\hat \rho_\calU(0)=\ket{1;\{0\}}\bra{1;\{0\}}$ [see Appendix]
\begin{align}\label{eq:Wfock11}
W_t(z)=\frac{e^{-\half z^T\cdot\calA^{-1}_0\cdot z}}{2\pi \sqrt{\det\calA_0}}\left(|u|^2z^T\cdot\calA^{-1}_0\cdot z-2|u|^2+1\right) \;,
\end{align}
which then, in the large-$t$ limit, [for $\eta>\eta_c$ note that $u(t) \to \mathcal Z e^{-i e_1 t}$] yields
\begin{equation}
W_t(z)=\frac{e^{-\half z^T\cdot\calA^{-1}_0\cdot z}}{2\pi \sqrt{\det\calA_0}}\left( c_0^4\ z^T\cdot\calA^{-1}_0\cdot z + 1 - 2c_0^4\right)\;.
\end{equation}
This is indeed the Wigner-function associated with Eq.~(\ref{eq:123}). 
On the other hand, if $\eta<\eta_c$, $u(t)\to 0$ in the long-time limit and $W_t(z)\to\frac{2}{\pi}e^{-2|z|^2}$, which is the Wigner function of the vacuum. 
Note that, while the number of total excitations is conserved, the single occupation number of the system $\hat a^\dagger \hat a$ is not. The RDM of the system can thus relax towards $\ket 0\bra 0$ for $\eta<\eta_c$. 

%
%

\vspace{1mm}
{\it Discussion and conclusion. -}
Let us briefly summarize what we have achieved in this letter. Eq.~(\ref{eq_Wf}) gives the general form of the Wigner function in the case of a Gaussian evolution, but \emph{arbitrary} initial conditions [in particular \emph{non-factorizing} ones] which are encoded in the Fourier transform of their Wigner function $\tilde W_0$. Since the time evolution [given after Eqs.~(\ref{eq_motion})] is in principle exactly solvable [note that we assumed $\hat H$ to be quadratic] a remaining difficulty may arise when calculating $\tilde W_0$ \emph{via} Eq.~(\ref{eq_def_W}). However, in the case where a superposition of Gaussian states is considered [i.e. Schr\"odinger cat states], or for Fock states, the effort for determining $\tilde W_0$ is minimal. 

In the case of a factorized initial  condition,  Eq.~(\ref{eq_factorized}) yields the evolution of the reduced initial Wigner function $W_\calS$ through the convolution with the propagator $K_t$. Eq.~(\ref{eq_factorized}) can be considered as the \emph{solution} to the corresponding \emph{master equations}: For instance it is straightforward to show that with $\hat \rho_\calB= e^{-\hat H_\calB/T}/Z_\calB$ [with $Z_\calB=\Tr\, e^{-\hat H_\calB/T}$], Eq.~\eqref{eq_factorized} reproduces the solutions of the exact master-equation of the QBM, see for instance \cite{Fleming2011}, and the resonant coupling \cite{Lei2011}, as shown in the Appendix. 

Finally, we have investigated the relaxation dynamics of the resonant model by analyzing its exact Wigner function. In contrast to previous interpretations, the so-called ``dissipationless'' dynamics, which appear when the $\calS\calB$-coupling is strong, are not caused by memory effects in the non-Markovian environment. It is rather a quantum phase transition in the whole system-bath ensemble which alters the dynamics. The emergence of the new ground state does not prevent the system from relaxing; on the contrary, we have shown that -- within the constraints imposed by energy and total particle conservation -- the system shows standard relaxation towards its new ground state. 
We want to stress that this peculiar behavior could not happen in the QBM, as the spectral function is unbounded in this model and therefore no stable pole can appear. 

We emphasize that our approach can be easily generalized to the case of several bosonic modes and used to study entanglement in such extended systems. Also, Eq.~(\ref{eq_factorized}) can be used as a guide to derive more easily the correct master equations for other systems in the future.

\emph{Acknowledgment. -} A.R. thanks F. Jendrzejewski for a suggestion and R. Blandino for enlightening discussions. J.B. thanks L. F. Cugliandolo for a critical reading of the manuscript. We acknowledge discussions with W. M. Zhang.

\appendix
\section{Appendix}
\subsection{Exact equation for the density-matrix matrix-elements}

We show here that the RDM, with Gaussian initial conditions, can be entirely written in terms of the non-equilibrium correlation functions. This can be demonstrated by starting from the average
\begin{align}
&\mean{e^{-i(y-x)\hat p(t)}e^{ir\hat q(t)}}_0 \equiv \mean{e^{-i(y-x)\hat p}e^{ir\hat q}}_t \nonumber\\
&\qquad=\Tr_\calU \Big\{e^{-i(y-x)\hat p}e^{ir\hat q}\hat \rho_\calU(t)\Big\} \nonumber\\
&\qquad=\int \rmd x'\bra{x'}e^{-i(y-x)\hat p}e^{ir\hat q}\hat \rho(t)\ket{x'} \nonumber\\
&\qquad=\int \rmd x' e^{ir(x'+x-y)}\rho_{x'+x-y,x'}(t) \; ,					
\label{eq_SM_rhoint}
\end{align}
where $\Tr_\calU\big\{\cdots\big\}= \int \rmd x'\bra{x'}\Tr_\calB\big\{\cdots\big\}\ket{x'}$.
If we multiply both sides of Eq.~\eqref{eq_SM_rhoint} by $\int\frac{\rmd r}{2\pi}e^{-i r x}$, and use the Baker-Campbell-Hausdorff formula, we find
\begin{equation}
\rho_{x,y}(t)=	\int\frac{\rmd r}{2\pi}e^{-\frac{i}{2}r(x+y)}\mean{e^{-i(y-x)\hat p(t)+ir\hat q(t)}}_0 \; .	
\label{eq_SM_rhoxy}
\end{equation}
The rhs of Eq.~\eqref{eq_SM_rhoxy} can be evaluated in a path integral formalism, where the operators $\hat q(t)$ and $\hat p(t)$ are replaced by c-numbers $q$ and $p$. Moreover, we know that $q$ and $p$ are, as soon as the initial condition is Gaussian, Gaussian random variables for all time. Therefore
\begin{equation}
\begin{split}
\rho_{x,y}(t)=	\int\frac{\rmd r}{2\pi}&e^{-\frac{i}{2}r(x+y-2\bar q)-i(x-y)\bar p} \\
&\times e^{-\half C_{pp}(x-y)^2 -\half C_{qq}r^2+r(x-y)C_{qp}} \; 	,
\label{eq_SM_rhoxy2}
\end{split}
\end{equation}
where $\bar A(t)=\mean{\hat A}_t$  and $C_{AB}(t)=\half\mean{\hat A\hat B+\hat B\hat A}_t-\bar A(t)\bar B(t)$ for any two operators $\hat A$ and $\hat B$. The integral over $r$ is then straightforward and one obtains Eq.~\eqref{eq_rhofin} of the main text.

\subsection{Equivalence with the solutions of the master equations}
\subsubsection{Case of the Quantum Brownian Motion}

We show here, that in the case of the QBM with factorized initial condition $\hat\rho_\calU(0)=\hat\rho_\calS\otimes e^{-\hat H_\calB/T}/Z_\calB$ and constant evolution Hamiltonian $\hat H$, our calculation  reduces to the known solution of the master equation, see for instance \cite{Fleming2011}. It is convenient to  write the interaction term $\hat H_{\rm QBM}=\sum_i g_i \hat q \ \hat q_i$ with $\hat q_i$ the position operator of the mode $i$. The renormalization of the frequency $\w_0$ can be absorbed in a redefinition as usual. With our notation, the solution of the master equation given in \cite{Fleming2011} is
\begin{equation}
\tilde W^{\rm ME}_t(k)= e^{-\half k^T\cdot\sigma\cdot k}\tilde W_\calS(\Phi^T(t)\cdot k) \; .
\end{equation}
where 
\begin{equation}
\sigma(t)= \int_0^t\rmd\tau\int_0^t\rmd\tau'\Phi(t-\tau)\cdot \begin{pmatrix}0&0\\0& \nu(\tau,\tau')\end{pmatrix}\cdot \Phi^T(t-\tau') \; ,
\end{equation}
and
\begin{equation}
\nu(\tau,\tau')=\sum_i \frac{g_i^2}{2m_i\w_i}\coth\Big(\frac{\w_i}{2T}\Big)\ \cos[\w_i(\tau-\tau')] \; .
\end{equation}
Our task is thus to show that in this case, $\tilde K_t(k)=e^{-\half k^T\cdot \sigma(t)\cdot k}$.

By definition, $\tilde K_t(k)=\tilde W_\calB(\{\calM_i^T(t)\cdot k\})$. If the density-matrix of $\calB$ is given by $e^{-\hat H_\calB/T}/Z_\calB$, it is factorized and therefore the initial Wigner function factorizes into a product over Wigner functions of each bath mode, $W_\calB({\zeta_i})=\prod_i W_i(\zeta_i)$, which in turn implies $\tilde K_t(k)=\prod_i\tilde W_i(\calM_i^T(t)\cdot k)$. Let us first focus on one mode $i$ of the bath. Its initial density-matrix is given by $e^{-\w_i \hat b^\dag_i\, \hat b_i/T}/Z_i$, the Wigner function of which is well known (see \emph{e.g.} \cite{LeonhardtBook})
\begin{equation}
W_i(\zeta_i)=\frac{e^{-\half \frac{\zeta_i^T\cdot \calA_i^{-1}\cdot \zeta_i}{1+2n(\w_i)}}}{2\pi \sqrt{\det \calA_i}\big(1+2n(\w_i)\big)} \; ,
\end{equation}
where $\calA_i$ is given in Eq.~\eqref{eq_SM_calA} and $n(\w_i)=1/(e^{\w_i/T}-1)$ is the Bose function. From this, we obtain
\begin{equation}
\tilde W_i(k)=e^{-\half \coth(\frac{\w_i}{2T}) \ k^T\cdot \calA_i\cdot k} \; ,
\end{equation}
where we have used $1+2n(\w_i)=\coth(\frac{\w_i}{2T})$. Next we need to find the expression of $\calM_i(t)$ [the definition of which is given in Eq.~\eqref{eq_Mi}] for the case under study. Here, the equations of motion are given by
\begin{equation}
\begin{split}
i\dt \hat z(t)&=i\begin{pmatrix} 0 & \frac{1}{m}\\ -m\w_0^2 & 0\end{pmatrix}\cdot \hat z(t)+i\sum_i  \begin{pmatrix} 0 & 0\\ g_i & 0\end{pmatrix}\cdot \hat \zeta_i(t) \; ,\\
i \dt \hat \zeta_i(t)& =  i\begin{pmatrix} 0 & \frac{1}{m_i}\\ -m_i\w_i^2 & 0\end{pmatrix}\cdot \hat \zeta_i(t)+i\begin{pmatrix} 0 & 0\\ g_i & 0\end{pmatrix} \cdot \hat z(t) \; ,
\end{split}
\end{equation}
From this, we find [one easily verifies that $\Phi$ solves the same equation as the corresponding one in \cite{Fleming2011}]
\begin{equation}
\begin{split}
\calL_i(t)&=\begin{pmatrix} \cos(\w_it) & \frac{\sin(\w_i t)}{m\w_i}\\ -m\w_i\sin(\w_i t) & \cos(\w_it)\end{pmatrix},\\
\calM_i(t)&=g_i\int_0^t \rmd\tau\ \Phi(t-\tau)\cdot\begin{pmatrix} 0 &0\\  \cos(\w_i\tau) & \frac{\sin(\w_i \tau)}{m\w_i}\end{pmatrix} \; .
\end{split}
\end{equation}
By using
\begin{equation}
\begin{split}
\begin{pmatrix} 0 &0\\  \cos(\w_i\tau) & \frac{\sin(\w_i \tau)}{m\w_i}\end{pmatrix}
\cdot\calA_i\cdot
&
\begin{pmatrix} 0 &  \cos(\w_i\tau') \\0& \frac{\sin(\w_i \tau')}{m\w_i}\end{pmatrix}
\\&=\begin{pmatrix} 0 &  0 \\0&\cos[\w_i (\tau-\tau')]\end{pmatrix},
\end{split}
\end{equation}
it is straightforward to obtain
\begin{equation}
\begin{split}
\tilde W_i(\calM_i^T(t)\cdot k)=e^{-\half k^T\cdot \sigma_i(t)\cdot k} \; ,
\end{split}
\end{equation}
with
\begin{equation}
\sigma_i(t)= \int_0^t\rmd\tau\int_0^t\rmd\tau'\Phi(t-\tau)\cdot \begin{pmatrix}0&0\\0& \nu_i(\tau,\tau')\end{pmatrix}\cdot \Phi^T(t-\tau') \; ,
\end{equation}
and 
\begin{equation}
\nu_i(\tau,\tau')=\frac{g_i^2}{2m_i\w_i}\coth\Big(\frac{\w_i}{2T}\Big)\ \cos[\w_i(\tau-\tau')] \; .
\end{equation}
We therefore have
\begin{equation}
\begin{split}
\tilde K_t(k)&=e^{-\half k^T\cdot \sum_i\sigma_i(t)\cdot k} \; ,\\
&=e^{-\half k^T\cdot \sigma(t)\cdot k} \; ,
\end{split}
\end{equation}
which had to be shown.


\subsubsection{Case of the resonant interaction}
We demonstrate here that the main formula~(\ref{eq_factorized}) yields also the solution of the exact master equation derived in the case of the resonant interaction $\hat H_r$ given in \cite{Lei2011}, with initial density-matrix $\hat\rho_\calU(0)=\hat\rho_\calS\otimes e^{-\hat H_\calB/T}/Z_\calB$ and constant Hamiltonian.

In the same spirit as in the previous calculation, let us start by determining $\tilde W_i(\calM_i^T(t)\cdot k)$ in the case of a resonant coupling, keeping in mind that $\tilde W_i(k)=e^{-\half \coth(\frac{\w_i}{2T}) \ k^T\cdot \calA_i\cdot k}$ as soon as the initial condition is given by $\hat\rho_\calS\otimes e^{-\hat H_\calB}/Z_\calB$.

In the resonant case, it is easier to solve the equations of motion for the ladder operators, which are
\begin{equation}
\begin{split}
i\dt\hat a(t)&=\w_0 \hat a(t)+\sum_i C_i \hat b_i(t) \; ,\\
i\dt\hat b_i(t)&=\w_i \hat b_i(t)+C_i \hat a(t) \; ,
\end{split}
\end{equation}
the solutions of which read
\begin{equation}
\begin{split}
\hat a(t)&=u(t)\hat a(0)+\int_0^t\rmd\tau \ u(t-\tau)\sum_i C_i e^{-i\w_i \tau}\hat b_i(0) \; ,\\
\hat b_i(t)&=e^{-i\w_it}\hat b_i(0)-iC_i \int_0^t\rmd\tau \ e^{-i\w_i(t-\tau)}\hat a(\tau) \; ,
\end{split}
\end{equation}
where the function $u(t)$ is such that $u(0)=1$ and $i\dt u(t)=\w_0 u(t)-i\int_0^t \rmd\tau \sum_i C_i^2 e^{-i\w_i(t-\tau)}u(\tau)$. The operator $\hat z$ is obtained by  $\hat z= \calP\cdot (\hat a,\hat a^\dag)^T$, with the transfer matrices
\begin{equation}
\begin{split}
\calP&= \begin{pmatrix}  \frac{1}{\sqrt{2m\w_0}} & \frac{1}{\sqrt{2m\w_0}}\\ -i \sqrt{\frac{m\w_0}{2}} & i \sqrt{\frac{m\w_0}{2}} \end{pmatrix} \; ,\\
\calP^{-1}&= \begin{pmatrix} \sqrt{\frac{m\w_0}{2}} & \frac{i}{\sqrt{2m\w_0}}\\ \sqrt{\frac{m\w_0}{2}} & \frac{-i}{\sqrt{2m\w_0}} \end{pmatrix} \; ,
\end{split}
\end{equation}
and a similarly for $\hat \zeta_i$, such that $\hat \zeta_i  = \calP_i\cdot (\hat b_i,\hat b_i^\dag)^T$. From this, we find that $\Phi(t)=\calP\cdot U(t)\cdot \calP^{-1}$, $\calL_i(t)=\calP_i\cdot L_i(t)\cdot \calP_i^{-1}$ and $\tilde M_{0i}=C_i \,\calP\cdot \calP_i^{-1}$, where
\begin{equation}
\begin{split}
U(t)&= \begin{pmatrix}  u(t) & 0\\ 0 & u^*(t)\end{pmatrix} \; ,\\
L_i(t)&= \begin{pmatrix}  e^{-i\w_i t} & 0\\ 0 & e^{i\w_i t} \end{pmatrix} \; .
\end{split}
\end{equation}
Furthermore, 
\begin{align}
&\calM_i\cdot \calA_i\cdot \calM_i^T = C_i^2 \calP\cdot\int_0^t\rmd\tau \int_0^t\rmd\tau' U(t-\tau)\cdot L_i(\tau) \nonumber
\\&\qquad\cdot \calP_i^{-1}\cdot \calA_i\cdot (P_i^{-1})^T\cdot L_i(\tau')\cdot U(t-\tau')\cdot\calP^T \; .
\end{align}
By using
\begin{equation}
\begin{split}
\calP_i^{-1}\cdot\calA_i\cdot(\calP_i^{-1})^T&= \begin{pmatrix}  0 & \half\\ \half & 0\end{pmatrix}
							=\calP^{-1}\cdot\calA_0\cdot(\calP^{-1})^T \; ,
\end{split}
\end{equation}
we finally obtain
\begin{align}
&\calM_i\cdot \calA_i\cdot \calM_i^T= \\
&\qquad\calA_0 \int_0^t\rmd\tau \int_0^t\rmd\tau' u(t-\tau)u^*(t-\tau') C_i^2 e^{-i\w_i(\tau-\tau')} \nonumber\; .
\end{align}
With the equal-time commutation relation of the ladder operators $[\hat a(t),\hat a^\dag(t)]=1$ [or equivalently by using the equation of motion of $u(t)$], one can show that
\begin{align}
 &\int_0^t\rmd\tau \int_0^t\rmd\tau' u(t-\tau)u^*(t-\tau') \sum_i C_i^2 e^{-i\w_i(\tau-\tau')}\nonumber\\
&\qquad=1-|u(t)|^2 \; .
\end{align}
After introducing $v(t)=\int_0^t\rmd\tau d\tau' u(t-\tau)u^*(t-\tau')\sum_i n(\w_i) C_i^2 e^{-i\w_i(\tau-\tau')}$ and by putting all pieces together, we obtain
\begin{equation}
\begin{split}
\tilde K_t(k)&=e^{-\half (1+2 v(t)-|u(t)|^2)k^T\cdot \calA_0\cdot k} \; ,\\
&=e^{-\half (1+2 v(t))k^T\cdot \calA_0\cdot k+\half\cdot k\cdot \Phi(t)\cdot \calA_0\cdot \Phi(t)^T\cdot k} \; ,
\end{split}
\end{equation}
where we have used $\Phi(t)\cdot \calA_0\cdot \Phi^T(t)=\calP\cdot U\cdot \calP^{-1}\cdot \calA_0\cdot (\calP^{-1})^T\cdot U\cdot \calP^T=|u(t)|^2\calA_0$. The second term in the exponential will compensate the opposite contribution which stems from $\tilde W_\calS(\Phi^T\cdot k)$. In order to compare this result with the one from the solution of the master equation, it is useful to express the Wigner function in terms of the initial $P$ function [$\tilde W_\calS(k)=\tilde P_\calS(k) \exp(-\half k^T\cdot \calA_0\cdot k)$]:
\begin{align}
\tilde W_t(k)&=e^{-\frac{1+2v}{2} k^T\cdot \calA_0\cdot k}\ \tilde P_\calS(\Phi^T\cdot k) \; ,\\
W_t(z)&=\int \rmd\tilde z\ \frac{e^{-\frac{1}{2(1+2v)}(z-\Phi\cdot \tilde z)^T\cdot \calA_0^{-1}\cdot (z-\Phi\cdot \tilde z)}}{2\pi \sqrt{\det\calA_0}(1+2v)}P_\calS(\tilde z) \; .
\label{eq_SM_sol_r}
\end{align}

The Wigner function solution of the master equation  is given in \cite{Lei2011} by (we have corrected here two typos \cite{note_SM})
\begin{equation}
W_t(\alpha)=\int \frac{\rmd^2\tilde\alpha}{\pi}\int \frac{\rmd^2\tilde\beta}{\pi}e^{-\frac{|\tilde\alpha|^2}{2}-\frac{|\tilde\beta|^2}{2}}\bra{\tilde\alpha}\hat\rho_0\ket{\tilde\beta} \calJ_t(\alpha;\tilde\alpha,\tilde\beta) \; ,
\label{eq_SM_WLei}
\end{equation}
where $\calJ_t$ is defined as
\begin{equation}
\calJ_t(\alpha;\tilde\alpha,\tilde\beta)=\frac{\Omega}{\pi}e^{-\Omega|\alpha|^2+\alpha^*\Omega u \tilde\alpha+\tilde\beta^*u^*\Omega \alpha+\tilde\beta^*(1-|u|^2\Omega)\tilde\alpha} \; ,
\end{equation}
with $\Omega=\frac{2}{1+2v}$. Here and below, the greek letters correspond the coherent state basis $z^T=\calP\cdot (\alpha,\alpha^*)^T$. We have also defined $\rmd^2\tilde\alpha=\rmd {\rm Re}( \tilde\alpha)\, \rmd{\rm Im}(\tilde\alpha)$ and this change of variables is such that the Jacobian is one.

The easiest way to show that Eq.~\eqref{eq_SM_WLei} is equivalent to our result is to write the initial density-matrix as $\hat \rho_\calS=\int \rmd^2\gamma\ P_\calS(\gamma)\ket\gamma\bra\gamma$ and perform the two Gaussian integrals over $\tilde\alpha$ and $\tilde\beta$, such that
$W_t(\alpha) = \int \rmd^2\gamma\ P_\calS(\gamma) I(\gamma)$ with
\begin{widetext}
\begin{align}
I(\gamma)&=\frac{\Omega}{\pi}\int \frac{\rmd^2\tilde\alpha}{\pi}\int \frac{\rmd^2\tilde\beta}{\pi}e^{-\frac{|\tilde\alpha|^2}{2}-\frac{|\tilde\beta|^2}{2}}\bra{\tilde\alpha}\gamma\rangle\langle\gamma\ket{\tilde\beta} e^{\alpha^*\Omega u \tilde\alpha+\tilde\beta^*u^*\Omega \alpha+\tilde\beta^*(1-|u|^2\Omega)\tilde\alpha-\Omega|\tilde\alpha|^2} \nonumber\\
&=\frac{\Omega}{\pi}\int \frac{\rmd^2\tilde\alpha}{\pi}\int \frac{\rmd^2\tilde\beta}{\pi}e^{-|\tilde\alpha|^2-|\tilde\beta|^2-|\gamma|^2+\tilde\alpha^*\gamma+\gamma^*\tilde\beta+\alpha^*\Omega u \alpha+\tilde\beta^*u^*\Omega \alpha+\tilde\beta^*(1-|u|^2\Omega)\tilde\alpha-\Omega|\tilde\alpha|^2} \nonumber\\
&=\frac{\Omega}{\pi}\int \frac{\rmd^2\tilde\alpha}{\pi}e^{-|\tilde\alpha|^2-|\tilde\gamma|^2+\tilde\alpha^*\tilde\gamma+\alpha^*\Omega u \tilde\alpha+\gamma^*u^*\Omega \alpha+\gamma^*(1-|u|^2\Omega)\tilde\alpha-\Omega|\tilde\alpha|^2} 
=e^{\alpha^*\Omega u \gamma+\gamma^*u^*\Omega \alpha-|u|^2\Omega|\gamma|^2-\Omega|\tilde\alpha|^2} \; ,
\end{align}
\end{widetext}
which then leads to
\begin{equation}
W_t(\alpha)=\int \rmd^2\gamma\ P_\calS(\gamma)\frac{\Omega}{\pi}e^{-\Omega|\alpha-u \gamma|^2 } \; .
\label{eq_SM_Wa}
\end{equation}
By using the following relations which allow to go back to the usual basis, $z^T=\calP\cdot (\alpha,\alpha^*)^T$ and $\tilde z^T=\calP\cdot (\gamma,\gamma^*)^T$,
\begin{equation}
\begin{split}
z-\Phi\cdot \tilde z &=\calP\cdot (\alpha-u\gamma,\alpha^*-u^*\gamma)^T \; ,\\
\half z^T\cdot \calA_0^{-1}\cdot z&=2|\alpha|^2 \; ,
\end{split}
\end{equation}
where the second equality comes from 
\begin{equation}
\calP^T\cdot \calA_0^{-1}\cdot \calP=\begin{pmatrix} 0 & 2\\ 2 & 0\end{pmatrix} \; ,
\end{equation}
one readily shows that Eq.~\eqref{eq_SM_Wa} is equal to Eq.~\eqref{eq_SM_sol_r}.

\subsection{Wigner function of a Fock state for the resonant model}

We compute now the Wigner function of the RDM of the initial factorized condition $\hat\rho_\calU(0)=\ket 1 \bra 1\otimes e^{-\hat H_\calB/T}/Z_\calB$. The Wigner function of the Fock state with one boson is well known and is given by
\begin{equation}
W_\calS(z)=\frac{z^T\cdot\calA^{-1}_0\cdot z-1}{2\pi \sqrt{\det\calA_0}}e^{-\half z^T\cdot\calA^{-1}_0\cdot z}\; ,
\end{equation}
the Fourier transform of which is 
\begin{equation}
\tilde W_\calS(k)=(1-k^T\cdot\calA_0\cdot k)e^{-\half k^T\cdot\calA_0\cdot k}\; .
\end{equation}
By using the formulae derived previously in the case of the resonant model, we obtain
\begin{equation}
\tilde W_t(k)=(1-|u|^2 k^T\cdot\calA_0\cdot k)e^{-(1+2v)\half k^T\cdot\calA_0\cdot k}\; ,
\end{equation}
and therefore
\begin{widetext}
\begin{equation}
W_t(z)=\left(|u|^2\ \frac{z^T\cdot\calA^{-1}_0\cdot z}{(1+2v)^2}-\frac{2|u|^2}{1+2v}+1\right)\frac{e^{-\half z^T\cdot\frac{\calA^{-1}_0}{1+2v}\cdot z}}{2\pi \sqrt{\det\calA_0}(1+2v)} \; ,
\end{equation}
which yields at $T=0$ [translating into $v\to 0$]
\begin{equation}
W_t(z)=\left(|u|^2z^T\cdot\calA^{-1}_0\cdot z-2|u|^2+1\right)\frac{e^{-\half z^T\cdot\calA^{-1}_0\cdot z}}{2\pi \sqrt{\det\calA_0}} \; .
\end{equation}
The Wigner function of an arbitrary Fock state $\ket n$ is
\begin{equation}
W_\calS(z)=\frac{(-1)^n L_n\big(z^T\cdot\calA_0^{-1}\cdot z\big)}{2\pi \sqrt{\det\calA_0}}e^{-\half z^T\cdot\calA^{-1}_0\cdot z}\; ,
\end{equation}
where $L_n(x)$ is the $n$-th Laguerre polynomial. It is not too hard to convince oneself that with this result, the Wigner function of the system after a time $t$ is given by 
\begin{equation}
W_t(z)=\left(1-\frac{2|u|^2}{1+2v}\right)^n L_n\left(\frac{|u|^2}{2|u|^2-1-2v}\frac{z^T\cdot\calA_0^{-1}\cdot z}{1+2v}\right)\frac{e^{-\half z^T\cdot\frac{\calA^{-1}_0}{1+2v}\cdot z}}{2\pi \sqrt{\det\calA_0}}\; .
\end{equation}
\end{widetext}

\vspace{-0.6cm}
\bibliographystyle{phjcp}

\begin{thebibliography}{10}
\bibitem{note0} Note that $C_{AB}(t)$ is an equal-time correlator and that $\mean{\hat A\, \hat B}_t=\mean{\hat A(t)\hat B(t)}_0$.
\bibitem{note1} It should be clear that the bath has already been integrated out in equation \eqref{eq_Wf}, as only \emph{one} integral is left (over the variable $k$) and not $N_\calB$. This last integral imposes the initial conditions.
\end{thebibliography}

\end{document}